\documentclass{article}
\usepackage{graphicx}
\usepackage{cite}
\usepackage{epsfig}
\usepackage{amsfonts}
\usepackage{amssymb}
\usepackage{amsmath}
\usepackage{slashed}

\usepackage{hyperref}
\usepackage{dcolumn}        
\usepackage{bm}            		 
\usepackage{amstext}
\usepackage[dvipsnames]{xcolor}
\usepackage{color} 
\usepackage{transparent}
\usepackage{rotating}
\usepackage{array}
\usepackage[T1]{fontenc}
\usepackage[utf8]{inputenc}
\newcommand\mybar{\kern1pt\rule[-\dp\strutbox]{.8pt}{\baselineskip}\kern1pt}
\usepackage{float}

\newcolumntype{P}[1]{>{\centering\arraybackslash}p{#1}}

\makeatletter
\newcommand{\vast}{\bBigg@{4}}
\newcommand{\Vast}{\bBigg@{5}}
\makeatother

\date{\today}

\begin{document}
\begin{center}

{\Large \bf A convenient gauge for virial identities in axial symmetry}
\vspace{0.8cm}
\\
{Jo\~ao M. S. Oliveira$^{1}$, 
Alexandre M. Pombo$^{2,3}$  \\
\vspace{0.3cm}
$^{1}${\small Centro de Matem\'atica, Universidade do Minho, 4710-057 Braga, Portugal}
\\
\vspace{0.3cm}
$^{2}${\small Departamento de Matem\'atica da Universidade de Aveiro and } \\ {\small  Centre for Research and Development  in Mathematics and Applications (CIDMA),} \\ {\small    Campus de Santiago, 3810-183 Aveiro, Portugal}
\\
\vspace{0.3cm}
$^{3}${\small Centro de Astrof\'\i sica e Gravita\c c\~ao - CENTRA,} \\ {\small Departamento de F\'\i sica,
Instituto Superior T\'ecnico - IST, Universidade de Lisboa - UL,} \\ {\small Avenida
Rovisco Pais 1, 1049-001 Lisboa, Portugal}
\vspace{0.3cm}
}
\end{center}

	\begin{abstract}
Virial identities are a useful mathematical tool in General Relativity. Not only have they been used as a numerical accuracy tool, but they have also played a significant role in establishing no-go and no-hair theorems while giving some physical insight into the considered system from an energy balance perspective. While the calculation of these identities tends to be a straightforward application of Derrick's scaling argument~\cite{derrick1964comments}, the complexity of the resulting identity is system dependent. In particular, the contribution of the Einstein-Hilbert action, due to the presence of second-order derivatives of the metric functions, becomes increasingly complex for generic metrics. Additionally, the Gibbons-Hawking-York term needs to be taken into account \cite{Herdeiro:2021teo}. Thankfully, since the gravitational action only depends on the metric, it is expected that a ``convenient'' gauge that trivializes the gravitational action contribution exists. While in spherical symmetry such a gauge is known (the $m-\sigma$ parametrization), such has not been found for axial symmetry. In this letter, we propose a ``convenient'' gauge for axial symmetry and use it to compute an identity for Kerr black holes with scalar hair.
	\end{abstract}

\medskip

%
%
\section{Introduction}\label{S1}
	In two previous papers, hereafter referred to as \textit{Paper I}~\cite{Herdeiro:2021teo} and \textit{Paper II}~\cite{herdeiro2022deconstructing}, a generalized framework of virial identities in field theory was presented. Such identities serve a variety of purposes, including establishing no-go theorems for both solitonic and black hole (BH) solutions (\textit{aka} no-hair theorems), as well as serving as guides to finding new solutions to the equations of motion and checking the accuracy of numerical solutions.

	As a brief summary, consider the following one-dimensional (1D) effective action (EA):
	\begin{equation}\label{actionspatial3}
	 \mathcal{S}^{\rm eff}[q_j(r),q'_j(r),q''_j(r),r] = \int_{r_i}^{\infty}\hat{\mathcal{L}}\left( q_j,q'_j,r\right)dr \ ,
	\end{equation}
	where $q_j$ are a set of $N$ parameterizing functions of some configuration, $j=1\dots N$, depending on a single ``radial'' coordinate $r$, $q_j'\equiv dq_j/dr$, $r_i$ is some appropriately chosen constant\footnote{With $r_i=0$ for solitons and $r_i=r_H$ for BHs, where $r_H$ is the event horizon radial coordinate.} and $\hat{\mathcal{L}}$ corresponds to the \textit{effective Lagrangian}. Where $\hat{\mathcal{L}}$ should depend on $q''_i$, the idea is that we can always absorb the second derivatives into a function $f$ as
	\begin{equation}
	 \hat{\mathcal{L}}\left(q_i,q'_i,q''_i,r\right)=\mathcal{L}\left(q_i,q'_i,r\right)+\frac{d}{dr}f\left(q_i,q'_i,r\right) \ ,
	\label{lplusf}
	\end{equation}
	with $f$ being some function that depends on the same variables as the non-total derivative piece of the effective Lagrangian, $\mathcal{L}$. Following Derrick's scaling argument \cite{derrick1964comments}, the scaling transformation $(\tilde{r}-r_i)\to \lambda (r-r_i)$ yields
	\begin{equation}
	 r\rightarrow \tilde{r} = r_i + \lambda(r-r_i) \ ,
	 \label{E3}
	\end{equation}
	which induces a variation of any fiducial configuration $q_j(r)$, as $q_j(r)\rightarrow q_{\lambda j}(r)=q_j( \tilde{r})$.
	The EA of the scaled configuration $q_{\lambda j}(r)$ becomes a \textit{function} of $\lambda$, denoted as $\mathcal{S}_\lambda^{\rm eff}$, and the true profile obeys the \textit{stationarity condition}
	\begin{equation}
	 \frac{\partial \mathcal{S}^{\rm eff}_\lambda}{\partial \lambda} \bigg|_{\lambda=1}=0 \ ,
\label{variationscale}
	\end{equation}
	the resulting identity is known as the virial identity
	\begin{equation}
	\int_{r_i}^\infty\left[ \sum_j \frac{\partial \mathcal{L}}{\partial q'_j} q'_j -\mathcal{L} -\frac{\partial \mathcal{L}}{\partial r}(r-r_i)\right]dr = \left[\frac{\partial f}{\partial r}(r-r_i) - \sum_i\frac{\partial f}{\partial q'_i} q'_i\right]^{+\infty}_{r_i} \ . 
\label{virialea4}
	\end{equation}

	For spherically symmetric configurations, \eqref{virialea4} can be readily applied to field theory models yielding their virial identity. In the case of General Relativity (GR), the fact that the Einstein-Hilbert (EH) action contains second derivatives of the metric, implies that the total derivative term in~\eqref{lplusf}, defined  by $f$, is non-zero. One of the key messages of Paper I was the importance of considering the Gibbons-Hawking-York (GHY)~\cite{york1972role,gibbons1993action} term as part of the gravitational action, to obtain the correct virial identity. Whereas this is certainly the case in general, there is a special ``gauge'' choice (by which we mean a choice of coordinates, plus a choice of metric functions) that simplifies the computation of the virial identity by trivializing the full gravitational contribution, $i.e.$ both EH action and the GHY term. In this case, the full virial identity is computed from an EA that comes completely from the ``matter'' (rather than gravitational) action. An example of such a gauge choice is what was called the $m-\sigma$ parametrization in Paper I:
	\begin{equation}\label{msigma}
	ds^2 = -\sigma^2\left(1-\frac{2m(r)}{r}\right)dt^2 + \frac{dr^2}{1-\frac{2m(r)}{r}} + r^2 \big(d\theta^2 + \sin^2\theta d\varphi^2\big) \ .
	\end{equation}
	For this case, the total derivative term contribution to the virial identity in \eqref{virialea4} cancels out with the Gibbons-Hawking-York term contribution, and we only need to consider the matter part of the effective action to calculate the identity. Assuming that $\mathcal{L}$ is composed of two terms, gravitational $\mathcal{L}_{grav}$ and matter $\mathcal{L}_{mat}$ terms, we have the following virial identity for the $m-\sigma$ ansatz
	\begin{equation}
	 \int_{r_i}^\infty\bigg[ \sum_j \frac{\partial \mathcal{L}_{mat}}{\partial q'_j} q'_j -\mathcal{L}_{mat} -\frac{\partial \mathcal{L}_{mat}}{\partial r}(r-r_i)\bigg]dr = 0 \ .
	\end{equation}
    As one might expect, a ``gauge'' choice like this significantly simplifies calculations, as the contribution of the gravitational part of the action tends to be very complicated. The simplicity associated with the lack of the gravitational part has allowed the use of these identities to establish no-hair~\cite{heusler1992scaling,heusler1996no,herdeiro2015asymptotically}, no-go theorems \cite{derrick1964comments} and to be widely used as a numerical accuracy tool \cite{herdeiro2017asymptotically,Blazquez-Salcedo:2020nhs,Fernandes:2019rez,Oliveira:2020dru,Fernandes:2019kmh,herdeiro2021imitation}. This is especially true for higher dimensional EAs, like stationary and axisymmetric spacetimes.

	In an attempt to tackle the latter, two requirements seem to be necessary: \textit{i)} the complete procedure to compute virial identities for a generic metric ansatz in GR and; \textit{ii)} a metric parametrization that trivializes the gravitational contribution. While the former was dealt with in Paper I and Paper II, the latter piece is still absent. 
	
	What we propose in this paper is the existence of a "convenient" gauge choice for stationary and axisymmetric spacetimes, which allows one to compute the complete virial identity from an EA that solely contains the "matter" action. This new gauge choice significantly simplifies both the computation and the overall identities, similar to the $m-\sigma$ gauge \eqref{msigma}. We hope this paper allows authors of future works to streamline the process of obtaining virial identities for more general spacetimes. A new mathematical tool that could be useful for a myriad of purposes, from establishing theorems to numerical accuracy.

    This paper is organized as follows. First we present the convenient gauge metric ansatz in Section \ref{S2}. Then, in the subsections \ref{S21} and \ref{S22}, we present two examples of the calculation of the virial identity: the electro-vacuum and Kerr black holes with scalar hair cases. Finally, in Section \ref{S3} we present our conclusions and a discussion. Throughout this paper we use units with $G=1=c$ and, to simplify notation, we use $\frac{d X}{dr}\equiv X'$ and $\frac{dX}{d\theta}\equiv \hat{X}$, with $X$ being a generic function that is $(r,\theta)-$dependent.
%
%
\section{An axially symmetric convenient gauge}\label{S2}
%
	The gravitational action contains two terms:
	\begin{equation}\label{Sgrav}
	 \mathcal{S}_{grav}=\mathcal{S}_{EH}+\mathcal{S}_{GHY}=\frac{1}{16\pi} \int_\mathcal{M} d^4x \sqrt{-g}R + \frac{1}{8\pi}\int_{\partial\mathcal{M}} d^3x \sqrt{-\gamma}(K-K_0) \ , 
	\end{equation}
	where $K =\nabla_\mu n^\mu$ is the extrinsic curvature of the boundary $\partial\mathcal{M}$ with normal $n^\mu$, and $\gamma$ is the associated 3-metric of the boundary. The extra $K_0$ term corresponds to the extrinsic curvature in flat spacetime (the background metric), necessary to obtain a finite result.

	The GHY boundary term is necessary for the well posedness of the gravitational action. The resulting virial identity can, in general, be expressed as

	\begin{equation}\label{grav}
	 \int _{r_i} ^{+ \infty} dr \int _0 ^\pi d\theta \ I_R = I_{GHY}\ .
	\end{equation}
	The standard metric ansatz used to describe a rotating gravitating object gives rise to a non-trivial contribution to the virial identities from the gravitational part of the action\footnote{A computation of the virial identity of the Kerr-Newman BH solution in Boyer-Lindquist and isotropic coordinates was performed and can be seen in Appendix~\ref{A}.}. As discussed in Paper I, for the spherically symmetric case there is a convenient gauge choice which trivializes the gravitational action contribution, making the virial identity originate purely from the matter action. In this section we discuss an ansatz for the axially symmetric case that produces the same effect: it trivializes the contribution from the gravitational part of the action.  

    Consider the following metric ansatz:
	\begin{equation}\label{Axialmetric}
	 ds^2 =-F_0^2 \, dt^2 +H(r) F_1^2\, dr^2 + (r-r_H) ^2  F_1^2\, d\theta ^2+F_2^2\Big(d\varphi - F_W\, dt\Big)^2 \ ,
\end{equation}	 
	where $\mathcal{F}_i\equiv \mathcal{F}_i(r,\theta)$ with $i=\{ 0,1,2,W\}$ are the four parameterizing functions and the radial coordinate $r\in [r_H,\, +\infty)$ while $r_i\equiv r_H$ corresponds to the radius of the horizon -- if it exist, which in these coordinates will always be a topological sphere \footnote{Observe that we are only interested in integrations outside the horizon $r>r_H$, where the metric is regular.}. Note that, for this metric ansatz, the radial scaling argument
	\begin{equation}\label{scaling}
	 r\to \tilde{r}=r_H+\lambda(r-r_H)\ .
	\end{equation}
	The remarkable property of ansatz \eqref{Axialmetric} is that, after scaling the metric functions \eqref{scaling} and respective derivatives $\big( \mathcal{F}_i'(r,\theta)\rightarrow \mathcal{F}_{i\, \lambda}'(r,\theta)/\lambda$ and $H'(r)\to H'_\lambda (r)/\lambda\big) $, it trivializes the gravitational contribution to the virial identity. 
	
	To show that, observe that the Ricci scalar in ansatz \eqref{Axialmetric} comes as:
	\begin{align}
	&\sqrt{-g}R =\frac{1}{2F_0 F_1 ^2 H^{3/2} (r-r_H)}\bigg\{ F_1 ^2 F_2 ^2 H \Big(H\hat{F}_W^2+(r-r_H)^2 F_W'^2\Big)+2F_0F_1 ^2 \bigg[-2H^2\partial_\theta(\hat{F}_0 F_2)\nonumber\\
	&+(r-r_H)\Big(-2\partial_r[F'_0 F_2 H (r-r_H)]+3 F'_0 F_2 H' (r-r_H)\Big)\bigg]+2F_0 ^2\bigg[ 2F_2 H \big(H\hat{F}_1^2+(r-r_H)F_1'^2\big)\nonumber\\
	&-2 H^2 F_1(\hat{\hat{F}}_1 F_2 + F_1 \hat{\hat{F}}_2) +(r-r_H)\bigg(\big[(r-r_H)H'-2H\big]F_1\partial_r(F_1 F_2)-2F_1 H(r-r_H)(F_2\hat{\hat{F}}_2+F_1 F''_2) + F_2 F_1^2 H'\bigg)\bigg]\bigg\}\ .
	\end{align}
	Note that, for each radial derivative $\mathcal{F}_i '$ there is an $(r-r_H)$ term accompanying -- with exception to the first $\frac{1}{r-r_H}$. Hence after scaling, the integral term associated with the Ricci scalar, $I_R$, becomes
	\begin{equation}\label{ZeroRicci}
	 \sqrt{-g_\lambda} R _\lambda = \frac{1}{\lambda}\sqrt{-g} R\ ,
	\end{equation}
	and therefore, the transformations $dr \rightarrow \lambda dr$ will keep the action invariant under scaling transformations. Thus, the EH action for this ansatz becomes scale invariant; the EH action does not contribute to the virial identity ($I_R =0$).
	
	Let us now show that the GHY boundary term also vanishes for this ansatz. In the metric parameterization \eqref{Axialmetric}, the normal vector is $n^\mu\partial_\mu= \frac{1}{\sqrt{H} F_1}\partial_r$ and $K_0=\frac{2}{(r-r_H)F_1}$, the extra total derivative from the GHY boundary term is
	\begin{align}
	 f^{GHY}=& \frac{F_0 F_2 (r-r_H)}{\sqrt{H}} \left[\frac{F_0'}{F_0} + \frac{F'_1}{F_1} + \frac{F'_2}{F_2} - \frac{1}{r-r_H}\Big(2\sqrt{H}-1\Big)\right]\ .
	\end{align}
	The resulting GHY contribution from $f^{GHY}$ to the virial identity is
	\begin{align}
	 I_{GHY}=\int _0 ^\pi d\theta \ 0\Big|_{r_H} ^\infty \ = 0\ .
	\end{align}
	Which is metric function independent. The GHY term is automatically null due to the form of the metric ansatz, meaning that this result is spacetime independent. The only remaining term of the action necessary to calculate the virial identity is the matter term. Note that, if we want to compute the virial identity for other ansatz\"e, one only needs to do a coordinate transformation after obtaining the identity using ansatz \eqref{Axialmetric} (see Appendix \ref{B} for an example).
%
\subsection{Examples: Electro-vacuum}\label{S21}
%
	The most generic BH solution containing both a spin and an electric charge is given by the Kerr-Newman family of solutions. In a generic coordinate system, the action that describes such an electro-vacuum solution is
	\begin{equation}
	\mathcal{S} = \mathcal{S}_{grav} + \mathcal{S}_{Maxwell}\ ,
	\end{equation}
	where $\mathcal{S}_{grav}$ is the gravitational action given by \eqref{Sgrav}, and $\mathcal{S}_{Maxwell}$ is the electromagnetic contribution:
	\begin{align}\label{Maxwell}
	 \mathcal{S}_{Maxwell}=-\int _{\mathcal{M}} d^4 x \ \frac{F_{\mu \nu}F^{\mu \nu}}{4}\ ,
	\end{align}
	Since we have already observed that the $\mathcal{S}_{grav}$ contribution to the virial identity is zero. The resulting virial identity is

	\begin{equation}\label{KNG2}
	 \int _{r_H} ^{+ \infty} dr \int _0 ^\pi d\theta\ I_M = 0\ ,
	\end{equation}
	where $I_M$ is the Maxwell contribution to the virial identity. The 4-vector potential ansatz comes as
	\begin{equation}\label{Amu}
	 A_\mu = \big(dt-a \sin ^2 \theta\, d\varphi\big) V(r,\theta)\ .
	\end{equation}
	The resulting Maxwell invariant computed with \eqref{Amu} and \eqref{Axialmetric} comes as
	\begin{align}
	 &\sqrt{-g} F_{\mu \nu}F^{\mu \nu} = 2 \frac{F_0F_2 \sqrt{H}}{r-r_H}\nonumber\\
	 &\times\bigg\{ -\frac{\hat{V}^2}{F_0 ^2}+a^2 \sin ^2 \theta\bigg[\bigg(\frac{1}{F_2 ^2}-\frac{F_W^2}{F_0^2}\bigg)\big(2\cos \theta V+\sin \theta \hat{V}\big)^2+\frac{(r-r_H)^2}{H}\bigg[a^2\sin ^4\theta \bigg(\frac{1}{F_2^2}-\frac{F_W^2}{F_0^2}\bigg)-\frac{1}{F_0^2}\bigg]V'^2\bigg\}\ .
	\end{align}
	Again, note that, with exception to the first $\frac{1}{r-r_H}$ term, each radial derivative of the vector function $V$ is accompanied by an $(r-r_H)$ term. Therefore, after implementing Derrick's scaling argument, the integral term associated with the Maxwell contribution, $I_M$, becomes
	\begin{equation}
	 \sqrt{-g_\lambda}\big( F_{\mu \nu} F^{\mu \nu}\big)_\lambda =\frac{\sqrt{-g}}{\lambda}\big(F_{\mu \nu} F^{\mu \nu}\big)
	\end{equation}
	and therefore the radial transformation $dr\to \lambda dr$ will keep the action invariant under the scaling transformation and the identity \eqref{KNG2} is trivially satisfied.
%
\subsection{Examples: Kerr BHs with scalar hair}\label{S22}
%
	As a specific example, let us consider a non-vacuum case where the Kerr BH can be embedded in a larger family of regular (on and outside an event horizon) BH solutions, namely, Kerr BHs with synchronised scalar hair~\cite{Herdeiro:2014goa,Herdeiro:2015gia}. The latter consists of a complex scalar field $\Phi$, minimally coupled to Einstein's gravity and under a self-interaction potential $U(|\Phi|)$, which can be seen as a generalization of spinning scalar boson star~\cite{herdeiro2019asymptotically,liebling2017dynamical,schunck2003general,schunck1998rotating}. Consider the action that describes a self-gravitating complex scalar field, described by the ansatz \eqref{Axialmetric} in a model with a a self-interaction potential $U(\Phi)$ 
	\begin{equation}\label{SEKBH}
	 \mathcal{S} = \mathcal{S}_{grav} -\frac{1}{4\pi} \int d^4 x \sqrt{-g} \bigg[\frac{1}{2}g^{\mu \nu}\big(\partial_\mu \Phi \partial_\nu \Phi^* +\partial_\mu \Phi^* \partial_\nu \Phi \big)+U(|\Phi|)\bigg]\ .
	\end{equation}
	Here $^*$ denotes complex conjugate. For the scalar field, one considers an ansatz with an harmonic azimuthal and time dependence while the scalar field amplitude $\phi$ is $(r,\theta)$-dependent
	\begin{equation}	 
	 \Phi(t,r,\theta,\varphi ) = \phi (r,\theta)e^{-i \omega t+i m \varphi}\ ,
	\end{equation}	
	with $\omega\in \mathbb{R}^+$ the frequency defining the harmonic time-dependence and $m\in\mathbb{Z}$ the azimuthal harmonic index. The resulting matter effective Lagrangian is
	\begin{equation}
	\mathcal{L} (\phi; \phi ', \hat{\phi};r,\theta) =(r-r_H) \sqrt{H}F_0 F_1^2 F_2 \bigg[\bigg(\frac{m^2}{F_2 ^2}-\frac{(\omega -m W)^2}{F_0 ^2}\bigg)\phi ^2+\frac{1}{F_1^2}\bigg(\frac{\phi'^2}{H}+\frac{\hat{\phi}^2}{(r-r_H)^2}\bigg)+U(\phi)\bigg]\ .
	\end{equation}
	
	Which, after applying Derrick's argument, gives the Virial identity
	\begin{equation}
	 \int _{r_H} ^{+\infty} dr \ \int _0 ^{\pi} d\theta \Big[ I_\Phi (\omega, m, r_H) + I_U ^{[\Phi] } (r_H) \Big] =0 \ ,
	\end{equation}
	where the virial terms are
	\begin{eqnarray}
	I_\Phi (\omega , m,r_H)  &=& \frac{2 (r-r_H) F_1^2 \sqrt{H}}{F_0 F_2} \bigg[ m^2 F_0 ^2-F_2 ^2  \big(\omega -m F_W ^2\big)\bigg] \phi^2 \ ,\\
	I_U ^{[\Phi]} (r_H) &=& 2 (r - r_H) F_0 F_1^2 F_2\ U(\phi) \ .
	\end{eqnarray}
	Note that, by setting $r_H=0$, one obtains the virial identity for the specific case of Boson Stars. To test these solutions we have used three Kerr BH with scalar hair configurations, all with a simple mass potential $U(\Phi)=\mu ^2 \Phi ^2 $; and two spinning scalar boson star configurations -- corresponding to the horizonless limit $r_H=0$ --, one with a simple mass potential $U(\phi)=\mu ^2 \phi ^2 $ and one with an axionic-like potential $U(\phi)=\frac{2 \alpha ^2}{B} \left(1-\sqrt{1-4 B \sin ^2\left(\phi/\alpha \right)}\right)$~\cite{delgado2020rotating,guerra2019axion}. In all cases, we obtained an absolute (relative) difference for the virial identity of $\sim 10 ^{-3}$ $(\sim 10^{-3})$.
	
		%
	\subsubsection{Stress-energy tensor decomposition}
%
	Following the work done in Paper II, we were able to reduce the virial identity of the previous subsection as a sum of the stress energy-tensor components. Since the gravitational part is zero, the virial identities reduce to
	\begin{equation}\label{E31}
	 \int d^3 x \sqrt{-g}\bigg\{ 2 \big(T_t ^t- T_r ^r -T_\theta ^\theta\big)+\big(T_t^t-T_\varphi ^\varphi \big)\bigg[1-\frac{4 \omega  F_2 ^2 (\omega -m F_W)}{m^2 F_0 ^2+F_2^2 \left(\omega ^2-m^2 F_W^2\right)} \bigg]\bigg\}=0\ ,
	\end{equation}
	which, unfortunately, has coefficients that are matter dependent, however it seems to support the preposition that virial identities can be decomposed in a sum of the stress-energy tensor terms.
	
	Expression \eqref{E31} also makes it so that the coordinate transformation of the virial identity to another metric is very simple. An example is given in Appendix \ref{B}.

%
\section{Conclusions and discussion}
\label{S3}
%
	In this work we have introduced an ansatz parametrization for axial symmetry that trivializes the gravitational contribution to the virial identity in GR. In particular, we show that in vacuum, the metric ansatz \eqref{Axialmetric} has a zero contribution from the Einstein-Hilbert term and from the Gibbons-Hawking-York boundary term. Concerning the latter, we observed that it is spacetime independent since the GHY term is immediately zero and does not depend on the asymptotics of the metric functions. 
	
	To summarize, the general method to calculate the virial identity for any metric using ansatz \eqref{Axialmetric} goes as follows:
	\begin{itemize}
	    \item Write the matter action $\mathcal{S}_{mat}$ using the convenient ansatz \eqref{Axialmetric}.
	    \item Apply the one-parameter radial scaling \eqref{scaling} while considering the one-parameter family of functions $\mathcal{F}_i^\lambda (r,\theta) = \mathcal{F}_i(\tilde{r},\theta)$ and $H^\lambda(r) = H(\tilde{r})$.
	    \item Apply Hamilton's principle by differentiating the resulting scaled action $\mathcal{S}_{mat}^\lambda$ with respect to $\lambda$ and set $\lambda=1$. The resulting virial identity is:
	    \begin{equation}\label{FinalVirial}
	        \frac{\delta \mathcal{S}_{mat}^\lambda}{\delta \lambda}\bigg|_{\lambda=1}=0 \ .
	    \end{equation}
	    \item To consider a different ansatz, perform a coordinate transformation of the terms in the virial integral identity \eqref{FinalVirial}. See Appendix \ref{B} as an example.
	\end{itemize}

    With the convenient metric ansatz, it was possible to compute the virial identity for a Kerr BH with scalar hair model. The latter becomes much simpler than the one obtained in Paper II due to the choice of metric parametrization. Concerning the latter example, a decomposition of the identity in a sum of the stress-energy tensor terms was also possible, supporting the proposition established in Paper II. However, this time, some of the parameters that multiply the stress-energy tensor terms are matter term-dependent, something not observed previously.
	
	At last, while we did not establish any new no-hair or no-go theorem with the aid of the new metric parametrization in this paper, the fact that the latter significantly simplifies the computation of the virial identities can seed new theorems in axial symmetry -- just like it was done in spherical symmetry through the use of the $m-\sigma$ parametrization. 
\appendix 
\addcontentsline{toc}{section}{APPENDICES}

%
	
\section*{Acknowledgements}
We would like to thank Carlos A. R. Herdeiro and Eugen Radu for the many valuable discussions. This work is supported by the Center for Research and Development in Mathematics and Applications (CIDMA) through the Portuguese Foundation for Science and Technology (FCT - Fundaç\~{a}o para a Ci\^{e}ncia e a Tecnologia), references UIDB/04106/2020, UIDP/04106/2020.  We acknowledge support from the projects PTDC/FIS-OUT/28407/2017, CERN/FISPAR/0027/2019, PTDC/FIS-AST/3041/2020 and CERN/FIS-PAR/0024/2021. This work has further been supported by the European Union’s Horizon 2020 research and innovation (RISE) programme H2020-MSCA-RISE-2017 Grant No. FunFiCO-777740.

\appendix

%
\section{Virial identities in axial symmetry}\label{A}
%
	The most generic BH solution containing both a spin and an electric charge is given by the Kerr-Newman family of solutions. In a generic coordinate system, the action that describes such an electro-vacuum solution is
	\begin{equation}
	\mathcal{S} = \mathcal{S}_{grav} + \mathcal{S}_{Maxwell}\ ,
	\end{equation}
	where $\mathcal{S}_{grav}$ is the gravitational action given by \eqref{Sgrav}, and $\mathcal{S}_{Maxwell}$ is the electromagnetic contribution given by \eqref{Maxwell}. For a generic metric ansatz, the resulting virial identity is
	\begin{equation}\label{KNG}
	 \int _{r_H} ^{+ \infty} dr \int _0 ^\pi d\theta \Big[I_R +I_M\Big] = I_{GHY}\ .
	\end{equation}
%
\subsection{Boyer-Lindquist coordinates}\label{A1}
%
	The Kerr-Newman metric can be expressed in the familiar Boyer-Lindquist (BL) coordinates. To compute the virial identity of a charged, spinning BH in BL coordinates, we will follow \cite{Townsend:1997ku}
	\begin{equation}\label{BL}
	ds^2 = -f_0 dt^2 + f_1 dr^2 + r^2 \big( f_3 d\theta ^2 +f_4 d\varphi ^2\big)-2rf_2 \, dt d\varphi\ ,
	\end{equation}
	where the $f_i$ functions, with $i\in \{0,1,2,3,4\}$, are all $(r,\theta)$-dependent.\footnote{Observe that the present metric is almost completely generic with the exception of the explicit radial dependence in the angular components. The latter seems to be a requirement for the finiteness of the virial identity contribution from the Ricci/GHY component.}
	The $4$-vector potential, $A_\mu$, ansatz 
	\begin{equation}
	 A_\mu = V (dt -a \sin ^2 \theta\, d\varphi)\ , \qquad \qquad {\rm with} \qquad V(r,\theta) = \frac{r\, Q}{\Sigma}\ ,
	\end{equation}
	with $Q$ the electrical charge. The metric functions $f_i$ in BL coordinates are given as
	\begin{align}
	f_0 & = \frac{\Delta - a ^2 \sin ^2 \theta}{\Sigma}\ , \qquad \qquad f_1 = \frac{\Sigma}{\Delta}\ ,\qquad\qquad f_2 = \frac{a}{r \Sigma} \Big[ \big(r^2 +a^2\big)-\Delta \Big]\sin ^2 \theta\ ,\nonumber\\
	f_3 & = \frac{\Sigma}{r^2 } \ ,\qquad \qquad f_4 =\frac{\sin ^2 \theta}{r^2 \Sigma}  \Big[ \big( a^2+r^2 \big) ^2-a^2 \Delta \sin ^2 \theta \Big]\ ,
	\end{align}
	with $\Delta = r^2 -2M r + a^2 + Q^2$ and $\Sigma = r^2 +a ^2 \cos ^2 \theta$.

	While the computation of the virial identity was done with the unknown metric functions $f_i$ the resulting expressions for the gravitating virial contribution are too complex to write. The resulting Einstein-Hilbert ($I_R$) and Maxwell ($I_M$) contributions in BL coordinates come as
	\begin{align}
	 I_R &= \frac{ r_H \sin \theta}{r^3 \big(\cos 2 \theta +1+2 \frac{r^2}{a^2}\big)^2}\bigg[ 12 \big( a^2+Q^2\big)-13 M r+39 r^2+ \Big(4 (a^2+Q^2)-3 M r+r^2\Big)\cos 4 \theta \nonumber\\
	 &+\frac{28 r^2}{a^2} \Big(\frac{4 Q^2 r^2}{7 a^2}- M r+Q^2+r^2\Big)+16 \Big( a^2-M r+Q^2+\frac{5}{2} r^2+\frac{r^2}{4a^2} \big(-9 M r+9 Q^2+5 r^2\big)\Big)\cos 2 \theta \bigg]\ ,\\
	 I_M &=\frac{Q^2 }{4r \Delta \Sigma ^6 \sqrt{\Omega}}\Bigg\{ \Omega  \sin ^2 \theta \Big[ \Delta \big( r^2-a^2 \cos ^2 \theta \big) \Omega \Big( a^2(3 r-2 r_H) \cos ^2 \theta-3 r^2 (r-2 r_H)\Big)-a^2 r \Big( 4 a^2 r \cos ^2 \theta (3 r-2 r_H) \Sigma ^2\nonumber\\
	 &+\Delta \sin ^2\theta \big( r^2-a^2 \cos ^2 \theta \big) \big( \Delta-a^2 \sin ^2\theta \big)\big( a^2 \cos ^2 \theta -r (r-4 r_H)\big)\Big)\Big]+4 a^2 r^3 (a^2+r^2)^2  \Sigma ^2 \big(\Delta -a^2 \sin ^2\theta \big)\cos ^2 \theta\Bigg\}\ ,
	\end{align}
	where we reintroduced the metric functions $\Sigma$ and $\Delta$ and added the function $\Omega=\left(a^2+r^2\right)^2-a^2 \Delta \sin ^2\theta $ to simplify the $I_M$ term. The GHY term, computed using the normal vector $n^\mu\partial_\mu= \frac{1}{\sqrt{f_1}}\partial_r$ and $K_0=\frac{2}{r \sqrt{f_3}}$, associated with these coordinates is
	\begin{equation}
	I_{GHY} = 4\frac{a^2 +Q^2}{r_H}\ .
	\end{equation}
%

%
\subsection{Isotropic coordinates}\label{A2}
%
	Consider the Kerr-Newman solution described by an axially symmetric metric gauge in isotropic coordinates
	\begin{equation}
	ds^2 = - \frac{f_1}{S} H ^2 dt^2+f_1\big( dr^2 + r^2 d\theta ^2\big)+\frac{r^2}{f_1} \sin ^2 \theta \Big( Wdt-S d\varphi\Big) ^2  \ ,
	\end{equation}
	where the metric functions $(f_0,f_1, W, S)$ are $(r,\theta)$-dependent and $H\equiv H(r)$. The $4$-vector potential comes as
	\begin{equation}
	A_\mu =\frac{V}{f_1}\Big[ dt -a \sin ^2 \theta\, d\varphi\Big]\ ,
	\end{equation}	
	while for the metric and Maxwell functions we have
	\begin{align}
	S & =\frac{1}{r^4}\Big[\Big(a^2+\big(-H r+M+2 r\big)^2\Big)^2-a^2 H^2 r^2 \sin ^2\theta\Big]\ ,\qquad \qquad 	f_1 = \frac{1}{r^2}\Big[ a^2 \cos ^2\theta +\Big(M-\big(H-2\big) r\Big)^2\Big]\ ,\\
	W &= \frac{a}{r^4} \Big[ a^2+(M+2 r) \Big(M-2 (H-1) r\Big)\Big]\ , \qquad \qquad H = 1+\frac{a^2-M^2+Q^2}{4 r^2}\ ,\qquad \qquad 	V  = \frac{Q}{r^2} \Big[ M-(H-2) r\Big]\ .
	\end{align}
	with
	\begin{align}
	I_R  =&\frac{\sin \theta}{2 H f_1 ^2 S^2} \Bigg\{ -2 S \Bigg[-H^2 f_1 \Big( r (r-2 r_H) f_1 ' S'+\hat{f}_1 \hat{S}\Big)+H f_1 ^2 \Big( H \big( (r-r_H) S'+\cot \theta \hat{S}\big)-r (r-2 r_H) H' S'\Big)\nonumber\\
	&+r^3W W' S' ( 3 r-4 r_H ) \sin ^2 \theta\Bigg]+S^2 \Bigg[ H \Bigg(2 f_1 \Big(H \big(-\hat{\hat{f}}_1+\hat{f}_1\cot \theta -r (r-2 r_H) f_1 ''\big)-r (r-2 r_H) H' f_1 \Big)\nonumber\\
	&+H\Big( \hat{f}_1 ^2+r (r-2r_H) f_1'^2\Big)+f_1^2 \Big(-4 r (r-2 r_H) H''-8 (r-r_H) H'\Big)\Bigg)+r^3 \sin ^2 \theta  (3 r-4 r_H) W'^2\Bigg]\nonumber\\
	&-H^2 f_1 ^2 \Big(r (r-2 r_H) S'^2+\hat{S}^2\Big)+r W^2 \sin ^2 \theta \Big( r^2 (3 r-4 r_H) S'^2+(3 r-2 r_H) \hat{S}^2\Big)\Bigg\}\ ,\\
	I_M = &-\frac{2 \sin \theta }{r^3 H f_1 ^5 S} \Bigg\{ -2 r^3 V f_1 \Bigg[r (r-2 r_H) V' f_1' \Big( S^2+a^2 W^2 \sin ^4\theta \Big)+2 a^2 V W^2 \sin ^3 \theta  \cos \theta \hat{f}_1\Bigg]\nonumber\\
	&+f_1 ^2 \Bigg[ a^2 V^2 \sin ^2 \theta \Big( H^2 \big(r^3 f_1 '^2+(r-2r_H) \hat{f}_1^2\big)+4 r^3 W^2 \cos ^2 \theta \Big)+r^4 (r-2r_H) V'^2 \Big( S^2+a^2 W^2 \sin ^4 \theta \Big)\Bigg]\nonumber\\
	&-2 a^2 V H^2 f_1^3 \Big( r^3 \sin \theta V' f_1 '+2 V \cos \theta (r-2 r_H) \hat{f}_1\Big)\sin \theta+r^3 V^2 \Big(\hat{f}_1^2+r (r-2 r_H) f_1 ^2\Big) \big( S^2+a^2 W^2 \sin ^4 \theta \big)\nonumber\\
	&+a^2 H^2 f_1 ^4 \Big( r^3 \sin ^2 \theta V'^2+4 V^2 \cos ^2\theta (r-2 r_H)\Big)\Bigg\}\ ,\\
	I_{GHY} &= 4(M-2r_H)\ .
	\end{align}
	Where the GHY term was computed using the normal vector $n^\mu\partial_\mu= \frac{1}{\sqrt{f_1}}\partial_r$ and $K_0=\frac{2}{\sqrt{f_1}r}$.
	
	For the Kerr-Newman BH in both BL and isotropic coordinates, we were unable to obtain a close form for the Einstein-Hilbert contribution. However, due to the relation \eqref{KNG} -- that has been numerically confirmed with arbitrary precision --, we know that the integral must give $I_{GHY}$.
	%
\section{Numerical metric transformation}\label{B}
%
	Let us consider the metric that we presented in Paper II, which is a widely used metric ansatz for numerical computation
	\begin{equation}\label{Eugenm}
	 ds^2 = -e^{2G_0}H(\bar{r}) dt^2+ e^{2G_1}\left[\frac{d\bar{r}^2}{H(\bar{r})}+\bar{r}^2d\theta^2\right] + e^{2G_2}\bar{r}^2\sin^2\theta\Big(d\varphi-G_Wdt\Big)^2 \ , \quad \;\;\; H(\bar{r})\equiv 1-\frac{r_H}{\bar{r}}\ ,
	\end{equation}
	where the parameterizing functions $\mathcal{G}_i$ with $i={0,1,2,W}$ are all $(\bar{r},\theta)-$dependent. The transformation between the convenient metric gauge \eqref{Axialmetric} and \eqref{Eugenm} comes as
	\begin{align}
	 & F_2 =e^{G_2} \bar{r} \sin \theta\ ,\qquad  F_1 = \frac{e^{G_1}}{H}\ ,\qquad F_W = G_W\ ,\qquad F_0 = e^{G_0}\ ,\qquad \bar{r} =r\ ,
	\end{align}
    With the new radial coordinate $r\in [r_H,\,+\infty)$. With this coordinate transformation, the virial identity for Kerr black holes with scalar hair is easily obtained from \eqref{E31} as
    \begin{equation}
	 \int d^3 x \sqrt{-g}\bigg\{ 2 \big(T_t ^t- T_r ^r -T_\theta ^\theta\big)+\big(T_t^t-T_\varphi ^\varphi \big)\bigg[1-\frac{4 \omega  e^{2G_2} \bar{r}^2 \sin^2 \theta (\omega -m G_W)}{m^2 e^{2G_0}+e^{2G_2} \bar{r}^2 \sin^2 \theta \left(\omega ^2-m^2 G_W^2\right)} \bigg]\bigg\}=0\ .
	\end{equation}
%
  \bibliographystyle{ieeetr}
  \bibliography{biblio}


\end{document}